\documentclass[12pt]{article}
\usepackage{graphicx}
\usepackage{indentfirst}
\usepackage{bm}

\begin{document}
\renewcommand{\topfraction}{1.}
\renewcommand{\bottomfraction}{1.}

\title{Magnetic moments and electromagnetic radii of nucleon and $\Delta(1232)$ in an extended GBE model}

\author{\begin{tabular}{c} \normalsize Jun He $^1$  and  Yu-Bing Dong $^{2,1}$ \\
\footnotesize \sl $^1$Institute of High Energy Physics, Chinese
Academy of Sciences,\\\footnotesize \sl P. O. Box 918-4, Beijing
100039, P. R. China
\\\footnotesize \sl$^2$Chinese Center of Advanced Science and Technology (World Laboratory)\\\footnotesize \sl
P. O. Box 8730, Beijing 100080, P. R. China \end{tabular}}

\date{\today}

\maketitle

\begin{abstract}
We derive the exchange currents of pseudoscalar, vector, and
scalar mesons from Feynman diagrams, and use them to calculate the
magnetic form factors of nucleon and $\Delta(1232)$. The magnetic
moments and electromagnetic radii are obtained by using those form
factors and the parameters determined from the masses of nucleon
and $\Delta(1232)$. We find the magnetic moments and
electromagnetic radii of nucleon and $\Delta(1232)$ can be
produced very well in the extended Goldstone-boson-exchange model
(GBE) in which all of pseudoscalar, vector and  scalar meson nonet
are included. The magnetic moments of $\Delta(1232)$ are closer to
experiment values and results from lattice calculation than the
results obtained by the model without other mesons except for pion
and sigma.
\end{abstract}
\normalsize

\section{Introduction}

In the frame of constitute quark model, the exchange currents
between quarks are important to produce the properties of hadrons.
There are two kinds of exchanges, gluon exchange and meson
exchange. In Isgur's model, one-gluon-exchange (OGE) governs the
structure of the hadrons\cite{IK}. Since the last few years, Shen
$et. al.$\cite{Shen}, Riska and Glozman \cite{GR, EGBE,G}applied
the quark-chiral coupling model to study the baryon structure. In
the works of Glozman $et.al.$ \cite{EGBE,G}, the vector meson
coupling was included to replace one-gluon-exchange. They found
the spin-flavor interaction is important in explaining the energy
of Roper resonance and got a comparatively good fit to the baryon
spectra. Recently, L. R. Dai $et.\ al.$ extended original chiral
SU(3) quark model, in which the nonet pseudoscalar meson exchanges
and nonet scalar meson exchanges are considered in describing
medium and long range part of the interaction and OGE contribute
the short repulse, to the extended chiral SU(3) quark model, in
which vector mesons are included\cite{Zhang}. They calculated N-N
scattering processes and found the similar results to the original
chiral SU(3) quark model calculation. In the $^1S_0$ case, the one
channel phase shifts of the extended chiral SU(3) quark model are
obviously improved. In our previous works \cite{PRD}, we found
that mixing angles of $N^*$ resonance are sensitive to the
different interaction models, and we calculated the amplitudes for
photoproduction of negative parity $N^*$ resonances under 2GeV to
compare different interaction models. All those suggest that
vector meson exchanges may be important to the short range
mechanism of quark-quark interaction.

On the other hand, the method,  by which the electromagnetic form
factors of deuteron are calculated on nucleon level \cite{Gari},
has been applied to quark level by Buchmann $et.\
al.$\cite{Buchmann1}. They have calculated the electromagnetic
property of nucleon and $\Delta$(1232)\cite{Buchmann2, Buchmann3}.
Their results of magnetic moments and electromagnetic radii of
nucleon agree with experiment values well while the magnetic
moment of $\Delta$(1232) is 6.981$\mu_N$.  There are two
experiment results, $6.14\pm0.51$$\mu_N$ by LOPEZCATRO and
$4.52\pm0.50\pm0.45$$\mu_N$ by BOSSHARD \cite{PDG}. Through the
result of Buchmann $et.al.$ is still within the region given by
Particle Data Group \cite{PDG}, it is larger than both experiment
values above. The recent result from lattice QCD is
$4.99\pm0.56$$\mu_N$, which supposes the latter experiment
result\cite{lattice}. In the calculations of Buchmann $et.al.$,
gluon, pion, scalar meson and confinement exchanges are
considered. But the effects of other pseudoscalar mesons and
vector mesons are not included. On nucleon level, the discrepancy
between the coupling constant of $\pi NN$ ($\approx14$) and that
of $\rho NN$ ($\approx3$) is large. The effect of $\rho$ meson is
negligible since there is square coupling constant in form
factors. However, by using the relations: $g_{\pi
qq}=\frac{3M_q}{5M_N}g_{\pi NN}$ and $g_{\rho qq}=g_{\rho
NN}$\cite{Riska}, we know that the coupling constant of $\pi qq$
is only about one fifth of $\pi NN$ while the coupling constant of
$\rho qq$ equals to that of $\rho NN$. Hence on the quark level,
the effect of $\rho$ meson on the form factors may not be ignored.
In this paper, we calculate magnetic moments and electromagnetic
radii of nucleon and $\Delta(1232)$ in the extended GBE model to
study the effects of the vector mesons.

In the following section, we present the Lagrangian we used. In
section 3, the potential  and masses of nucleon and $\Delta(1232)$
in the extended GBE will be calculated to determine the
parameters. The magnetic moments and electromagnetic radii will be
obtained in section 4. Conclusion and discussion will be given in
the last section.

\section{Lagrangian used}

First, we give the Lagrangians which will be used to derive
potentials and the exchange currents. The total Lagrangian can be
written as following:
\begin{eqnarray}\label{langrangiantotal}
&&{\cal
L}_{total}=\frac{i}{2}(\overline{\psi}\gamma^\mu\partial_\mu\psi-\partial_\mu\overline{\psi}\gamma^\mu\psi)-m_q\overline{\psi}
\psi+\sum_{\pi,\eta,\eta'}{\cal
L}_\varphi+\sum_{\rho,\omega,\phi}{\cal L}_V+{\cal L}_\sigma .
\end{eqnarray}
where $\psi$ and $m_q$ are the quark field and quark mass. Because
we only calculate the case of nucleon and $\Delta$(1232), we do
not consider strange meson exchanges.

Under a global infinitesimal chiral transformation, we can obtain
the $\pi$ part of Lagrangian as following:
\begin{eqnarray}
&&{\cal L}_{\pi}=-g_{\pi
q}\overline{\psi}i\sum_{a=1}^3\lambda^a\cdot
\pi^a\gamma_5\psi+\frac{1}{2}\sum_{a=1}^3D_\mu\pi^a\cdot
D^\mu\pi^a,\ {\rm where:}\ \ D\pi^a=\partial_\mu
\pi^a-g_\rho\sum_{b,c=1}^3\epsilon_{abc}\rho_\mu^b\pi^c.\hspace{10pt}\
\end{eqnarray}
Here and henceforth $\pi^a$ and $\rho_\mu^a$ are the pion and rho
meson fields, respectively,  $\lambda^a_i$ is the Gell-mann flavor
matrix of $i$-th quark which reduce to Pauli matrix ${\bm \tau}_i$
as a=1,2,3 because we do not consider the strange dimension.
$g_{\pi q}$ and $g_\rho$ are coupling constants.

We construct the $\rho$ quark coupling as Jido $et.\ al.$ did in
the case of $\rho$ nucleon coupling \cite{Oset}. According to the
construction of the gauge theory, the direct coupling of a quark
$\psi$ to the $\rho$ meson is constructed replacing the derivative
by the associated covariant derivative with the gauge symmetry:
\begin{eqnarray}
&&\partial_\mu \psi(x)\rightarrow D_\mu \psi(x)=\partial_\mu
\psi(x)+ig_\rho\sum_a[\psi(x),V^a(x)]\rho_\mu^a,
\end{eqnarray}
where $V^a$ is the generator of the gauge symmetry of the SU(3)
flavor space. Here, $[\psi(x),V^a(x)]=-\frac{1}{2}\lambda^a\psi$.
Then the $\rho$ part of Lagrangian is as following:
\begin{eqnarray}
&&{\cal L}_{\rho}=-g_{\rho
q}\overline{\psi}\gamma^\mu\sum_{a=1}^3\lambda^a\cdot
\rho^a_\mu\psi-\frac{1}{4}\sum_{a=1}^3\rho^{\mu\nu a}\cdot
\rho^a_{\mu\nu},\  {\rm where:}\ \ \rho_{\mu\nu}=\partial_\mu
\rho^a_\nu-\partial_\nu
\rho^a_\mu-g_\rho\sum_{b,c=1}^3\epsilon_{abc}\rho^b_\mu\rho^c_\nu,\hspace{15pt}
\end{eqnarray}
where we assume coupling constant between rho and quark, $g_{\rho
q}=\frac{g_\rho}{2}$, in order to be consistent to the $\pi$ part
in form.

For $\eta$ and $\omega$ mesons, Lagrangians have  analogous forms
as the part of $\pi$ and $\rho$, respectively.

We also consider $\sigma$ exchange :
\begin{eqnarray}
&&{\cal L}_{\sigma}=g_{\sigma
q}\overline{\psi}\sigma\psi+\frac{1}{2}(\partial\sigma)^2.
\end{eqnarray}

The electromagnetic Lagrangians can be derived using the principle
of minimal coupling $\partial_\mu\rightarrow\partial_\mu+iqA_\mu$,
where q is the charge carried by the field upon which the
derivative operator acts. Then the $\gamma q\overline{q}$,
$\gamma\pi\pi$ and $\gamma\rho\rho$ interaction Lagrangians are:
\begin{eqnarray}
 {\cal L}_{\gamma
q\overline{q}}&=&-Qe\overline{\psi}\gamma_\mu\psi
A^\mu,\label{qqq}\\
{\cal L}_{\gamma
\pi\pi}&=&-ie[\pi\partial_\mu\pi^+-\pi^+\partial_\mu\pi]A_\mu,\label{qqqq}\\
{\cal L}_{\gamma
\rho\rho}&=&-ie[\partial^\mu\rho_\nu^+(A_\mu\rho^\nu-A^\nu\rho_\mu)-\partial^\mu\rho_\nu(A^\nu\rho_\mu^+-A_\mu\rho^{\nu+})],
\end{eqnarray}
where
\begin{eqnarray}
\pi=\frac{1}{\sqrt{2}}[\pi^1+i\pi^2],\ \
\rho_\mu=\frac{1}{\sqrt{2}}[\rho_\mu^1+i\rho_\mu^2],
\end{eqnarray}
and $Q$ is the charge of quark in unit of $e$. $A_\mu$ is the
electromagnetic field. Because the charges of $\pi^0,\rho^0,
\omega,\phi,\eta, \eta' {\rm and}\ \sigma$ are zero, there is no
interaction between them and photon.

\section{The extended GBE model}

In the harmonic-oscillator model, we assume that the Hamiltonian
of three quark system is of the form:
\begin{equation}\label{Hamiltonian}
\label{Ham} H=\sum_{i=1}^{3} \Bigl ( m_q+ {{\bf p}^{2}_{i}\over
2m_{q}} \Bigr ) + \sum_{i<j}^{3} V^{conf}({\bf r}_i,{\bf r}_j) +
\sum_{i<j}^{3} V^{res}({\bf r}_i,{\bf r}_j),
\end{equation}
where ${\bm r}_i$, ${\bm p}_i$ are the spatial and momentum
coordinates of the $i$th quark, respectively. The third term is a
two-body harmonic-oscillator confinement potential:
\begin{equation}
\label{conf} V^{conf}({\bf r}_i,{\bf r}_j)= -a_C
\bm{\lambda}^c_i\cdot \bm{\lambda}_j^c ({\bf r }_i-{\bf r }_j)^2,
\end{equation}
 and $a_C$
presents the strength of the confinement and $\bm{\lambda}^c_i$ is
the color operator of the i-th quark. After removing the kinetic
energy of the center of mass motion of three quark system, we take
the first three terms of Eq. (\ref{Hamiltonian}) as unperturbed
hamiltonian $H_0$. Hence the total baryon wave function
$\Phi_{N(\Delta)}$ is an inner product of the orbital,
spin-isospin, and color wave function and given by
\begin{eqnarray}
\label{ostcwf} \mid \Phi_{N (\Delta)}>= ({ \pi b^2})^{-3/2}
\exp(-(\bm{\rho}^2+\bm{\lambda}^2)/2b^2)) \mid ST>^{N(\Delta)}
\times \mid [111]>^{N(\Delta)}_{color},
\end{eqnarray}
where the Jacobi coordinates $\bm{\rho}$ and $\bm{\lambda}$ are
defined as $\bm{\rho}=({\bf r}_1-{\bf r}_2)/\sqrt{2}$ and
$\bm{\lambda}=({\bf r}_1+{\bf r}_2-2{\bf r}_3)/\sqrt{6}$.

The residual interaction $V^{res}$ consists gloun and meson
(pseudoscalar mesons $\pi$, $\eta$, $\eta'$, vector mesons $\rho$,
$\omega$, $\phi$, and scalar meson $\sigma$) exchange potentials.
In the following, we will explain them in detail.

The one-gluon exchange potential\cite{deR75}:
\begin{eqnarray}
\label{gluon} V_{g} (ij) & = & {\alpha_{s}\over 4}
\bm{\lambda}_i^c\cdot\bm{\lambda}_j^c \Biggl \lbrace {1\over r}-
{\pi\over m_q^2} \left ( 1+{2\over 3}
\bm{\sigma}_{i}\cdot\bm{\sigma}_{j} \right ) \delta({\bf r})
-{1\over 4m_q^2} {1\over r^3} \left ( 3\bm{\sigma}_i\cdot  {\hat
{\bf r} } \, \bm{\sigma}_j\cdot{\hat{\bf r}} -
\bm{\sigma}_i\cdot\bm{\sigma}_j \right ) \nonumber \\ & & -{1\over
2m_q^2} {1\over r^3} \biggl [ 3 \left ( {\bf r} \times {1\over2}
({\bf p}_i-{\bf p}_j) \right ) \cdot
{1\over2}(\bm{\sigma}_i+\bm{\sigma}_j)  - \left ( {\bf r}\times
{1\over2} ({\bf p}_i+{\bf p}_j) \right ) \cdot
{1\over2}(\bm{\sigma}_i-\bm{\sigma}_j) \biggr ] \Biggr \rbrace,
\hspace{15pt}\
\end{eqnarray}
where ${\bf r}={\bf r}_i-{\bf r}_j$; $\bm{\sigma}_i$ is the usual Pauli
spin matrix.

The one-$\pi$ exchange potential:
\begin{eqnarray}
&&V_\pi(ij)= \sum_{a=1}^3\lambda_i^a\cdot \lambda_j^a{\bm
\sigma}_i\cdot{\bm \nabla}{\bm \sigma}_j\cdot{\bm
\nabla}f_\pi(r),\end{eqnarray} where
\begin{eqnarray}
f_M(r)=\frac{g_{M q}^2}{4\pi(2m_q)^2}\left(\frac{e^{-\mu_M
r}}{r}-\frac{e^{-\Lambda_M r}}{r}\right),
\end{eqnarray}
$M$ presents $\pi$, $\eta$, $\eta'$, $\rho$, $\omega$ or $\phi$
meson. $\mu_M$, $\Lambda_M$ and $g_{M q}$ is meson mass, cut-off
and meson quark coupling constant.

The one-$\eta$($\eta'$) exchange potential can be obtained from
one-$\pi$ exchange potential through replacing the all things
about $\pi$ by corresponded ones of $\eta$.

The one-$\rho$ exchange potential:
\begin{eqnarray}
&&V^0_\rho(ij)=\sum_{a=1}^3\lambda_i^a\cdot \lambda_j^a{\bm
\sigma}_i\times{\bm \nabla}\cdot{\bm \sigma}_j\times{\bm
\nabla}f_\rho(r),\\ &&V^{LS}_\rho(ij)=
\sum_{a=1}^3\lambda_i^a\cdot \lambda_j^a2[ 3{\bm
r}\times\frac{1}{2}({\bm p}_1-{\bm p}_2)\cdot{1\over2}({\bm
\sigma}_1+{\bm \sigma}_2)-{\bm r}\times\frac{1}{2}({\bm p}_1+{\bm
p}_2)\cdot{1\over2}({\bm
\sigma}_1-{\bm \sigma}_2)]f_{\rho}(r),\hspace{15pt}\ \\
&&V^{C}_\rho(ij)= \sum_{a=1}^3\lambda_i^a\cdot
\lambda_j^a4m_q^2f_{\rho}(r).
\end{eqnarray}

The one-$\omega$ and one-$\phi$ exchange potentials can be
obtained from one-$\rho$ exchange potential by analogous
replacement.

Here, we use the "perfect" mixing pattern for the $\eta$, $\eta'$,
$\omega$ and $\phi$, $i. \ e. $,
\begin{eqnarray}
&&\lambda_\eta=\frac{1+\sqrt{2}}{\sqrt{6}}\lambda_8+\frac{-1+\sqrt{2}}{\sqrt{6}}\lambda_0,
\ \
\lambda_{\eta'}=-\frac{-1+\sqrt{2}}{\sqrt{6}}\lambda_8+\frac{1+\sqrt{2}}{\sqrt{6}}\lambda_0;\label{flavoreta}\\
&&\lambda_\omega=\sqrt{\frac{1}{3}}\lambda_8+\sqrt{\frac{2}{3}}\lambda_0,
\ \ \ \ \ \ \ \ \ \ \ \ \ \ \
\lambda_{\phi}=-\sqrt{\frac{2}{3}}\lambda_8+\sqrt{\frac{1}{3}}\lambda_0.\label{flavoreta1}
\end{eqnarray}

Hence we only consider nucleon, the strange dimension in the
Gell-mann matrices above have no effect on the results. Then Eqs.
(\ref{flavoreta},\ref{flavoreta1}) become:

\begin{eqnarray}
\lambda_\eta\rightarrow\frac{1}{\sqrt{2}}{\bf 1}, \ \
\lambda_{\eta'}=\frac{1}{\sqrt{2}}{\bf 1};\ \ \lambda_\omega={\bf
1}, \ \ \lambda_{\phi}=0;
\end{eqnarray}
that is, there is no contribution from $\phi$.

The scalar meson exchange potential is as following:
\begin{eqnarray}
V_{\sigma}(ij) & = & -{g_{\sigma q}^2\over {4 \pi}}  \left (
{e^{-\mu_{\sigma} r}\over r}- {e^{-\Lambda r}\over r} \right ).
\end{eqnarray}

With the Hamiltonian of Eq.(\ref{Ham}) and the wave function of
Eq.(\ref{ostcwf}) it is straightforward to calculate the nucleon
mass. One obtains
\begin{eqnarray}
\label{massN} M_{N}(b)& = & 3m_q+{3\over 2m_q b^2} + V_{conf}(b)
-2\alpha_s \sqrt{{2\over \pi}}{1\over b}
+\frac{1}{4}\delta_g  -\frac{5}{4}\delta_\pi+\frac{1}{4}\delta_\eta-\frac{5}{2}\delta_\rho+\frac{1}{2}\delta_\omega+V_{\sigma}(b),\hspace{15pt}\  \\
M_{\Delta}(b)&=& 3m_q+{3\over 2m_q b^2} + V_{conf}(b) -2\alpha_s
\sqrt{{2\over \pi}}{1\over b} +\frac{5}{4}\delta_g
-\frac{1}{4}\delta_\pi-\frac{1}{4}\delta_\eta-\frac{1}{2}\delta_\rho-\frac{1}{2}\delta_\omega+V_{\sigma}(b),\hspace{15pt}\
\label{massD}
\end{eqnarray}

where
\begin{eqnarray}
 \delta_g (b)&=& {4 \alpha_s  \over 3
\sqrt{2\pi} m_q^2  b^3 }, \\ \label{deltapi} \delta_{M}(b)&= &-4
 {f^2_{\pi q} \over {4 \pi}
\mu_M^2} \sqrt{2\over \pi}{1\over b} \Bigl \{ \mu_M^2 \Bigl (
1-\sqrt{\pi}({\mu_M b\over \sqrt{2}})e^{\mu_M^2b^2/2} {\rm
erfc}({\mu_M b \over \sqrt{2}}) \Bigr )-(\mu_M \leftrightarrow
\Lambda_M) \Bigr \},
\end{eqnarray}
where the individual terms in Eqs. (\ref{massN},\ref{massD}) are
the nonrelativistic  kinetic energy, quadratic confinement, gluon,
pion, eta, rho, omega and sigma  contributions, respectively. In
this paper we use $\eta$ to present both $\eta$ and $\eta'$. Here
$\delta_\eta$ presents $\frac{1}{2}(\delta_\eta+\delta_{\eta'})$.
The confinement contribution to the nucleon and $\Delta$ mass is
given by
\begin{eqnarray}
V_{conf}(b)=24 a_c b^2,
\end{eqnarray}
and the $\sigma$-meson potential contribution is
\begin{eqnarray}
\label{Vsigma} V_{\sigma}(b)=-6 {g^2_{\sigma q}\over 4 \pi} {1
\over  \sqrt{2\pi}}{1\over b} \Bigl \{ \Bigl (
1-\sqrt{\pi}({\mu_{\sigma} b\over \sqrt{2}})
e^{\mu_{\sigma}^2b^2/2} {\rm erfc}({\mu_{\sigma} b \over
\sqrt{2}}) \Bigr )- (\mu_{\sigma} \leftrightarrow \Lambda) \Bigr
\}.
\end{eqnarray}

In our calculation, we leave out all spin-orbit forces and the
central components from the vector meson exchanges as Ref.
\cite{EGBE}. This treatment is further supported on more
theoretical grounds by a study of the two-pion exchange mechanism
between constituent quarks in Ref. \cite{twopion}.

Subtracting Eq.(\ref{massN}) from Eq.(\ref{massD}) all
spin-independent terms drop out and one gets
\begin{eqnarray}
\label{M-M}
M_{\Delta}-M_N= \delta_g(b)+ \delta_{\pi}(b)-\frac{1}{2}\delta_\eta+2\delta_\rho-\delta_\omega,
\end{eqnarray}

To obtain the numerical results, we show the parameters in our
calculation. First we assume constituent quark mass
$m_q=\frac{1}{3}M_N=313MeV$. The gluon quark coupling constant
$\alpha_s$ is determined by Eq. (\ref{M-M}). Here we use the same
values of meson quark coupling constants for pseudoscalar and
vector mesons respectively as Ref. \cite{EGBE}, which is also used
in Ref. \cite{Zhang}. The coupling constants between vector mesons
and quark in Ref. \cite{EGBE} have included the part of tensor
coupling. We follow them ont only in calculating mass but also in
calculating magnetic moments and electromagnetic radii because the
effect of tensor coupling part can be included by replacing vector
coupling constants by the sum of vector and tensor coupling
constant. This can be seen by comparing the exchange currents
below of Gari \cite{Gari} and ours. The values of Cut-offs
$\Lambda_\gamma$ follow the linear scaling laws\cite{EGBE}:
\begin{eqnarray} \Lambda_\gamma\ =\ \Lambda_\pi\ +\ \kappa\ (\mu_\gamma\ -\ \mu_\pi\ )   &&for\ pesudoscalar\  mesons\ , \\
 \Lambda_\gamma\ =\ \Lambda_\rho\ +\ \kappa\ (\mu_\gamma\ -\ \mu_\rho\ ) &&for\ vector\  and\  scalar\  mesons\ .\end{eqnarray}
The parameters b and $a_c$ are determined to get the right mass of
nucleon. The explicit values of parameters are presented in Table
\ref{paramter}.

\begin{table*}[h]
\caption{\small Parameters: The quark mass is assumed to be one
third of nucleon mass. The masses of mesons are from Particle Data
group\cite{PDG}. The cut-offs, quark size and the coupling
constants between mesons and quark are from Ref. \cite{Buchmann3}.
b and $a_c$ are determined to get the right mass of
nucleon.}\label{paramter}
\renewcommand\tabcolsep{0.21cm}\begin{tabular}{ c  c c  c c c c c}  \hline\hline
$m_q$               &  313Mev       &  $\mu_\pi$                       &   139Mev                          &    $\mu_\eta$              &  547Mev               &    $\mu_{\eta'}$      &  958Mev  \\
$\mu_\rho $         &  770Mev       &  $\mu_{\omega}$                  &   782Mev                          &    $\mu_{\sigma}$          &  680Mev  \\
$g_{Ps q}^2/4\pi$   &  0.67         &$(g^v_{V q}+g^t_{V q})^2/4\pi$    &  1.31                             &$g_{\sigma}^2/4\pi$         &  0.67                 \\
$\Lambda_\pi$       &700MeV         &$\Lambda_\rho$                    &1200MeV                            &$\kappa$                    &1.2                    \\
 \hline
$\alpha_s$          &0.8584          &b                                &0.609fm                            &$a_c$                       & 17.98MeVfm$^{-2}$\\
\hline\hline
\end{tabular}
\end{table*}

The contributions from each exchange potential is listed in Table
\ref{mass}. $\alpha_s$ here is smaller than Ref. \cite{Buchmann2},
1.093. This is consistent with the conclusion of Ref.\cite{Zhang}
that the strength of OGE is reduced after the addition of vector
mesons. This result a gap of contribution from gluon between this
paper and Ref. \cite{Buchmann2} about 120MeV. This gap is
cancelled by the contribution from rho meson exchange. The
contribution of $\omega$-exchange is small. Then the mass of
nucleon is reproduced with a small adjustments of b and $a_c$.

\begin{table*}[h]
\caption[Nucleon mass]{\small Contributions of the kinetic energy
(without the rest mass term) and individual potential terms in the
Hamiltonian to the nucleon mass of Eq.(\ref{massN}) $cc$: color
Coulomb part of $V_{g}$, $\delta$: $\delta$-function part of
$V_{g}$. All entries are in [MeV]. }\label{mass}
\renewcommand\tabcolsep{0.3cm}\begin{tabular}[h]{r|r rr r r r r r r r} \hline\hline
& $ T^{kin}$ & $V^{conf}$ & $V^{g}_{cc}$ & $V^{g}_{\delta}$ &
$V^{\pi}$ &$V^{\eta}$ &$V^{\rho}$ &$V^{\omega}$ &
$V^{\sigma}$ & Total \\
\hline
N       &503.1 & 160.0 & -443.9 & 39.6 & -127.2&9.5&-126.7&24.8 & -39.4& 0.0  \\
$\Delta$&503.1 & 160.0 & -443.9 & 198.2& -25.4 &-9.5&-25.3&-24.8 & -39.4& 293.0  \\
\hline\hline
\end{tabular}
\end{table*}

\section{Magnetic form factors, moments and radii of
nucleon}

By using the Lagrangians above, we can evaluate Feynman diagrams,
which are shown by Figs. A1(a-e) in Appendix A, as Refs.
\cite{Buchmann1, Buchmann2, Buchmann3,Dubach}. The total current
operator consists of the usual one-body operator and two-body
exchange current operators tightly related to the different
quark-quark interactions:
\begin{eqnarray}\label{density}
&&\rho^{}_{total}({\bf q}) =\sum_{i=1}^3 {\rho}^{}_{imp}({\bf
r}_i)+ \sum_{i<j}^3 \{  \rho^{}_{g q\bar q}( {\bf r}_i, {\bf r}_j)
 +\sum_{M=\pi, \eta, \eta',\rho,\omega}\rho_{Mq\bar q}( {\bf r}_i, {\bf
 r}_j)+\sum_{S=\sigma,conf}\rho_{Sq\bar q}( {\bf r}_i, {\bf
 r}_j)\},\hspace{15pt}\
\end{eqnarray}
\begin{eqnarray}\label{current}
&&{\bf J}^{}_{total}({\bf q}) = \sum_{i=1}^3 {\bf J}^{}_{imp}({\bf
r}_i)\nonumber\\
& +& \sum_{i<j}^3 \{  {\bf J}^{}_{g q\bar q}( {\bf r}_i, {\bf
r}_j)
 +\sum_{M=\pi, \eta, \eta',\rho,\omega}{\bf J}_{Mq\bar q}( {\bf r}_i, {\bf r}_j)
+\sum_{M=\pi,\rho}{\bf J}_{\gamma MM} ( {\bf r}_i, {\bf
 r}_j)+\sum_{S=\sigma,conf}{\bf J}_{Sq\bar q}( {\bf r}_i, {\bf
 r}_j).\hspace{15pt}\
\end{eqnarray}

The explicit form of each current is given in Appendix A.

Quite generally, the charge radius is defined as the slope of the charge
form factor at zero-momentum transfer
\begin{eqnarray}
\label{chr}
r^2_C= -{6 \over F_C(0)} {d \over d { {\bf q}}^2 } F_{C}({\bf q}^2)\mid_{{\bf q}^2=0},
\end{eqnarray}
where, according to the general definition of the elastic form
factors \cite{Gari},
\begin{eqnarray}
\label{chaff}
F_C({\bf q}^2)= {\sqrt{4\pi}} < J M_J\! \!=\! \!J \, \, T M_T \mid
         {1\over 4\pi} \int d\Omega_q \rho({\bf q}) Y^0_0(\hat{\bf q})
         \mid J M_J\!\!=\!\!J \,\, T M_T>.
\end{eqnarray}

We also consider the constituent quark size. It is an effective
measure to including relativistic effects\cite{Buchmann3,Stanley}.
We use $\langle r^2\rangle_q=0.36 fm^2$\cite{Buchmann3}, which is
consistent to Ref. \cite{Krutov}.

Using Eq.(\ref{density}) and the ground state wave functions of
Eq.(\ref{ostcwf}), we obtain charge radius of $\Delta^+$ as
following:
\begin{eqnarray}
\label{rdel} r^2_{\Delta^+} = b^2 + r_{\gamma q}^2+ {b^2\over 6
m_q} (5\delta_g - \delta_{\pi}- \delta_{\eta}-2
\delta_{\rho}-2\delta_{\omega}) + {5 \over 6 m_q^3} V_{conf }+
r_{\sigma}^2.
\end{eqnarray}

The charge radius of $\Delta^0$ is zero exactly, which is
consistent with the analysis in Ref. \cite{Buchmann2}.

If we compare this with the corresponding result for the proton
\begin{eqnarray}
\label{rpmec} r^2_{p}  =  b^2 + r_{\gamma q}^2+ {b^2\over 2 m_q}
(\delta_g - \delta_{\pi}-2\delta_{\rho}) + {5 \over 6 m_q^3}
V_{conf }+ r_{\sigma}^2,
\end{eqnarray}
and neutron
\begin{eqnarray}
\label{rnmec}
r^2_{n}  =
-{b^2\over 3m_q} (\delta_g+ \delta_{\pi}-\frac{1}{2}\delta_{\eta}+2\delta_{\rho}-\delta_{\omega}) = -b^2 {M_{\Delta}- M_N
\over M_N},
\end{eqnarray}
we obtain from Eqs.(\ref{rdel}-\ref{rnmec}) the
parameter-independent result
\begin{eqnarray}
r^2_{\Delta}=r^2_p-r^2_n.
\end{eqnarray}

So we find that the relations Eqs. (38,39) found in Ref.
\cite{Buchmann2} are still tenable through we consider both
pseudoscalar and vector mesons.

We list our numerical results in Table \ref{Charg radii}. The
results agree with the experiment values well. By using Eq.
(\ref{rnmec}) and experiment values, we can determined b=0.610fm,
which agrees with the value b=0.609fm that we obtain in section 3
and use in the calculation.

\begin{table*}[h]
\caption{\small Nucleon and $\Delta$(1232) charge radii from
individual two-body exchange currents. A finite electromagnetic
quark size $r^2_{\gamma q}=0.36$ fm$^2$ is used. The charge radius
of the $\Delta^{0}$ is zero in the present model. The experiment
values are from Particle Data Group\cite{PDG}. All entries are in
[fm$^2$] except for the total result which is in [fm].
}\label{Charg radii}
\renewcommand\tabcolsep{0.185cm}\begin{tabular}[htb]{ r | r  r rrr r  r r  cr }\hline\hline
          & $r^2_{imp}$& $r^2_{gq\bar q}$ & $ r^2_{\pi q\bar q}$ & $ r^2_{\eta q\bar q}$ & $ r^2_{\rho q\bar q}$ & $ r^2_{\omega q\bar q}$ & $ r^2_{\sigma}$ &$r^2_{conf}$& $ \sqrt{ \mid r^2_{total} \mid} $& Exp.\\ \hline
$p $      & 0.731    & $ 0.094 $ & $ -0.060 $   & $0.000$       & $-0.060 $     &0.000            &0.029            &$-0.169$  &   0.751                     & 0.870                \\
$ n $     & $ 0.000$ & $-0.063 $ & $ -0.040$    & $ 0.007$      & $ 0.040 $     & $0.020 $        & 0.000           &0.000     & 0.340                        &0.341\\
$\Delta $ & 0.731    & $ 0.157 $ & $ -0.020 $   &$-0.002$       &$-0.020$       & $-0.010 $       &0.029            &$-0.169$  &  0.830\\ \hline \hline
\end{tabular}
\end{table*}

The magnetic moments of $\Delta$ are defined as the $q \to 0$
limit of the magnetic dipole form factor \cite{Gari}
\begin{eqnarray}
\label{magff}
F_M({\bf q}^2)= {2 \sqrt{6\pi} M_N \over iq} < J M_J\!\!=\!\!J \, \, T M_T\mid
         {-i\over 4\pi} \int d\Omega_q
         [ Y^1(\hat{\bf q}) \times {\bf J} ({\bf q}) ]^1
         \mid J  M_J\!\!=\!\!J \, \,T M_T>,
\end{eqnarray}

We calculate nucleon and $\Delta$(1232) magnetic moments, and
$N\to \Delta$ transition magnetic moments. We consider
next-to-leading order term of the isovector pion pair-current not
only in the case of $\Delta$(1232) but also in the case of
nucleon, and find it is really no obvious effect on the result.
The numerical results are presented in Table \ref{Magnetic
moments}. It's found the magnetic moments of nucleon agree with
experiment very well. The magnetic moments of $\Delta$ are
proportional to their charge as only gluon, pion is considered.
There are obvious improvements for magnetic moments of $\Delta$.
For $\Delta^{++}$, the effect of addition of $\eta$ and vector
mesons gives a decrease about 1.3 from the values of Buchmann
$et.\ al.$, 6.9814$\mu_N$\cite{Buchmann2}. This value is more
consistent with the experiment values, $6.14\pm0.51$ by LOPEZCATRO
and $4.52\pm0.50\pm0.45$ by BOSSHARD \cite{PDG}. It is just in the
margin of the recent result from lattice QCD, $4.99\pm0.56$
\cite{lattice}. The values of magnetic moment for $\Delta^+$ we
obtained agree with the lattice values, $2.49\pm0.47\mu_N$
\cite{lattice}, and the experiment values,
$2.7^{+1.0}_{-1.3}(stat.)\pm1.5(syst.)\pm3(theor)$
\cite{kotulla}compared with the value, 3.491, in Ref.
\cite{Buchmann2}.

There is only a small improvement for the $N\to \Delta$ transition
magnetic moments. Our result is consistent with some authors,
2.44$\mu_N$ by simple nonrelativistic quark model
(NRQM)\cite{PDG}, 2.53(2.63)$\mu_N$ by Dahiya $et.\
al.$\cite{Dahiya} and $1.7\sim3.0\mu_N$ by Julia-Diaz $et.\ al.$
\cite{Julia-Diaz}. The experiment values adopted by Dahiya $et.\
al.$ is 3.1 $\mu_N$\cite{PDG96}. The recent one is
$3.642\pm0.019\pm0.085\mu_N$\cite{NDexp}. The results of NRQM,
Dahiya $et.\ al.$, Julia-Diaz $et.\ al.$ and ours are smaller than
the experiment values especially the latter. The problem with the
underestimation of the $N \to \Delta$ transition magnetic moment
persists also after inclusion of both pseudoscalar and vector
meson exchange currents.

\begin{table*}[h]
\caption{\small Nucleon and $\Delta$(1232) magnetic moments, and
$N\to \Delta$ transition magnetic moments from individual two-body
exchange currents. The experiment values of nucleon, $\Delta$ and
N-$\Delta$ transition are $\mu_p=2.792847337(29) \mu_N$,
$\mu_n=-1.91304272(45)\mu_N$, $\mu_{\Delta^{++}}=6.14\pm0.51\mu_N$
(by LOPEZCATRO) and $4.52\pm0.50\pm0.45$ (by BOSSHARD)\cite{PDG},
$\mu_{\Delta^{+}}=2.7^{+1.0}_{-1.3}(stat.)\pm1.5(syst.)\pm3(theor)$
\cite{kotulla} and $\mu_{p\to
\Delta^{+}}=3.642\pm0.019\pm0.085\mu_N$\cite{NDexp}, respectively
.
 All entries are in $\mu_N$.  }\label{Magnetic moments}

\renewcommand\tabcolsep{0.045cm}\begin{tabular}[h]{ l | r  r r  r  r  r  rrrrrr } \hline\hline
                    & $\bm{\mu}_{imp}$ & $\bm{\mu}_{gq\bar q}$ & $\bm{\mu}_{\pi q\bar q}$ &$\bm{\mu}_{\gamma \pi \pi}$ & $\bm{\mu}_{\eta q\bar q}$& $\bm{\mu}_{\rho q\bar q}$ &$\bm{\mu}_{\gamma \rho \rho}$ & $\bm{\mu}_{\omega q\bar q}$& $\bm{\mu}_{\sigma}$ & $\bm{\mu}_{conf}$ &$\bm{\mu}_{total} $          \\ \hline \hline
$p $                & $3.000 $     & $ 0.473 $      & $ -0.141 $               & $ 0.469$                   &0.000                     &0.000                      &$-0.051$                      & $-0.148$                   &$0.252$              &$-1.023 $     & $ 2.830 $              \\
$ n $               & $ -2.000$    & $-0.158 $      & $0.195  $                & $-0.469$                   & 0.020                    &$-0.101$                   &0.051                         &0.049                       &$-0.168$             & $0.682 $     & $-1.897 $       \\ \hline
$ \Delta^{++} $     & $ 6.000 $    & $1.891$        & $0.325 $                 & $0.000 $                   & $ 0.121$                 &$-0.604$                   & $0.000$                      &$-0.592$                    &$0.504$              & $-2.045$     & $5.599 $                    \\
$ \Delta^{+} $      & $ 3.000 $   & $0.945$       & $0.162 $                & $0.000 $                   &$0.061$                  &$-0.302$                      &0.000                         &$-0.296$                       &$0.252$             & $-1.023$      & $2.799 $                    \\
$ \Delta^{0} $      & $ 0.000 $    & $0.000 $       & $0.000 $                 & $0.000 $                   & $ 0.000$                 & $ 0.000$                  & $ 0.000 $                    &0.000                       &0.000                &0.000         &0.000                          \\
$ \Delta^{-} $      & $ -3.000 $   & $-0.945$       & $-0.162 $                & $0.000 $                   &$-0.061$                  &0.302                      &0.000                         &0.296                       &$-0.252$             & $1.023$      & $-2.799 $                  \\ \hline
$p\to \Delta^{+}$   & $ 2.828 $    & $0.223$        & $-0.276 $                & $0.663 $                   & $-0.057$                 &0.142                      &$-0.073$                      & $-0.140$                    & $0.237 $            &$-0.964$      &2.585 \\
\hline \hline
\end{tabular}
\end{table*}

The magnetic radius is defined as the slope of the magnetic form
factor at zero momentum transfer:
\begin{eqnarray}
\label{magr} r^2_M= -{6 \over F_M(0)} {d \over d { {\bf q}}^2 }
F_{M}({\bf q}^2)\mid_{{\bf q}^2=0},
\end{eqnarray}

we list our results of magnetic radii of the nucleon and
$\Delta$(1232) in Table \ref{Magnetic radii}. The results agree
with experiment very well.
\begin{table*}[h]
\caption {\small Magnetic radii of the nucleon and $\Delta$(1232)
from individual two-body exchange currents. The magnetic radius of
the $\Delta^0$ is zero. A finite electromagnetic quark size,
$r^2_{\gamma q}=0.36$ fm$^2$, is used. Experiment values is from
Refs. \cite{Sick, Kubon, Hammer}. All entries are in [fm$^2$],
except for total results which are in [fm].
%
}\label{Magnetic radii}
\renewcommand\tabcolsep{0.11cm}\begin{tabular}[h]{ r | r r  r  r  r  r  rr rrrr} \hline\hline
           & $r^2_i$    & $r^2_{gq\bar q}$ & $ r^2_{\pi q \bar q}$ & $r^2_{\gamma \pi \pi }$ & $ r^2_{\eta q \bar q}$ & $ r^2_{\rho q \bar q}$ & $r^2_{\eta \rho \rho }$ & $ r^2_{\omega q \bar q}$ & $ r^2_{\sigma}$ & $r^2_{c}$ & $ \sqrt{ \mid r^2_{t} \mid} $ &Exp.\\ \hline
$p $       & $0.775 $   & $ 0.094$  & $-0.029 $             &
$0.217 $                & $0.000 $               &$0.000 $
& $-0.011 $               &$-0.011$                  & $0.047$
& $-0.331$  & 0.867                         & 0.855    \\ \hline $
n $      & $ 0.771 $  & $0.047 $  & $-0.263 $             & $0.324
$                & $-0.006 $              & 0.011
&$-0.016$                 &$-0.005$                  &  $0.047 $
& $-0.329 $ & $0.885$                       & 0.873 \\ \hline $
\Delta $ & $ 0.783 $  & $0.191$   & $0.034 $              & $0.000
$                & $0.011$                & $0.220$
&$0.000$                  &$-0.022$                  & $0.048$
&$ -0.335 $ & $0.830 $                      &     \\ \hline \hline
\end{tabular}
\end{table*}

From Tables 3-5, we can find the effects of mesons except for pion
and sigma is non-negligible. They provide a contribution to give
better result for magnetic moment of $\Delta$. In other cases the
result is similar to the one without those mesons since the effect
of those mesons can be smeared by the variation of parameters.

\section{Conclusion and discussion}
We calculate magnetic moments and electromagnetic radii of nucleon
and $\Delta(1232)$ in an extended GBE model. With the parameters
determined by the mass of nucleon and $\Delta(1232)$, the magnetic
moments and electromagnetic radii of nucleon are consistent with
the experiment results very well. There are obvious improvements
for magnetic moments of $\Delta$. Our result is more consistent
with experiment results and  results from lattice QCD calculation
compared with the model which only consider pion and sigma mesons.
The relations obtained from one-pion exchange survive from
addition of other mesons. The results above should be helpful in
better understanding the short range mechanism of quark-quark
interaction.

\Large\begin{flushleft} {\bf
Acknowledgements}\end{flushleft}\normalsize

This work is supported by the National Natural Science Foundation
of China (No. 10075056 and No. 90103020), by CAS Knowledge
Innovation Project No. KC2-SW-N02.

\renewcommand\theequation{A\arabic{equation}}
\renewcommand\thefigure{A\arabic{figure}}
\setcounter{equation} 0 \setcounter{figure} 0

\Large\begin{flushleft} {\bf Appendix A: Charge density and
exchange current }\end{flushleft}\normalsize

By using the Lagrangian in Eq. (1), we can evaluate Figs. A1(a-e)
to obtain charge density and exchange currents as Refs.
\cite{Buchmann1, Buchmann2, Buchmann3,Dubach}.We obtain the charge
density and exchange current as follows:

\begin{figure*}[h]
  \includegraphics[bb=0 450 180 800,clip=true,scale=0.34]{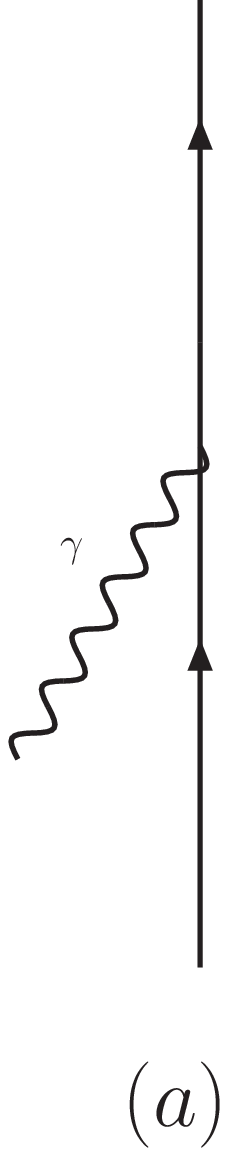}
  \includegraphics[bb=80 450 330 800,clip=true,scale=0.34]{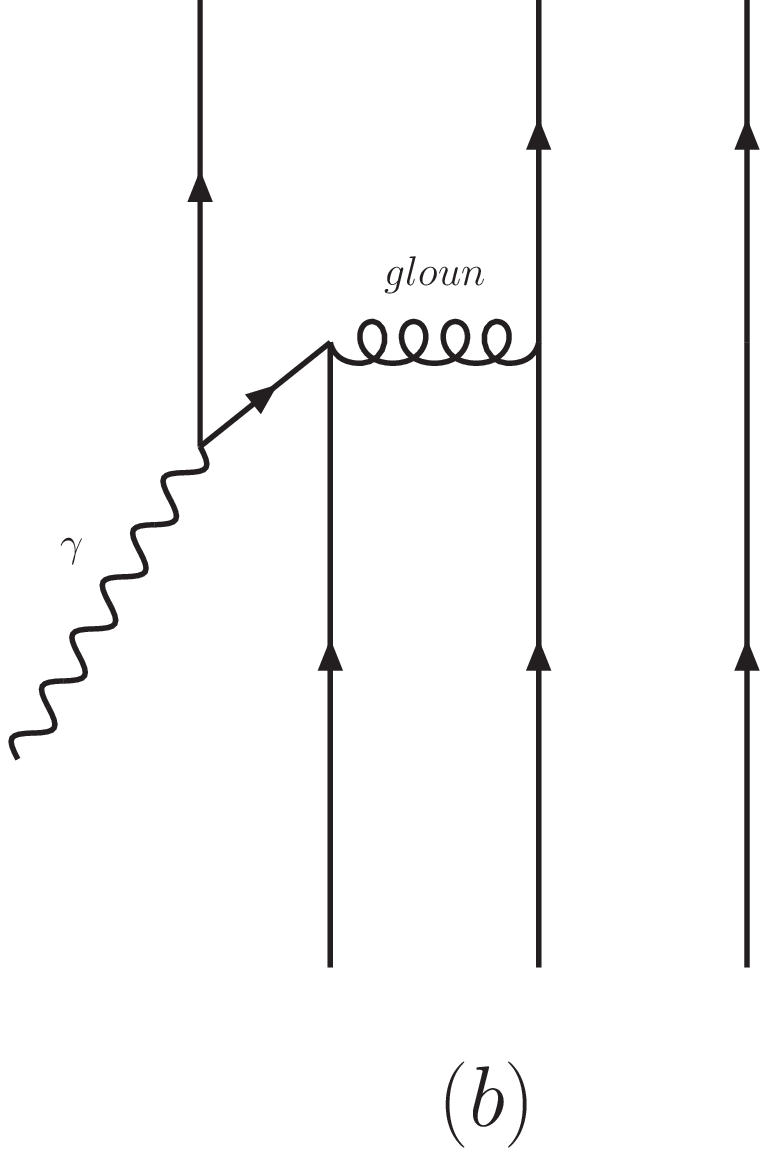}
  \includegraphics[bb=80 450 330 800,clip=true,scale=0.34]{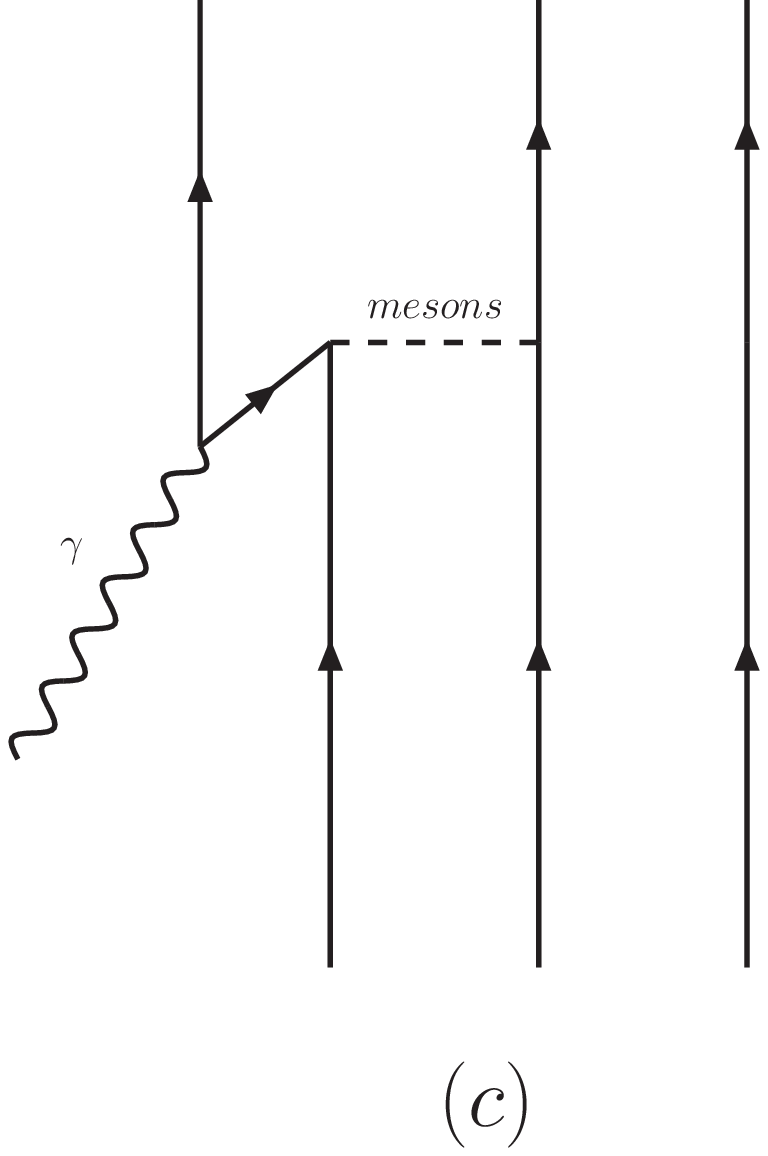}
  \includegraphics[bb=80 450 330 800,clip=true,scale=0.34]{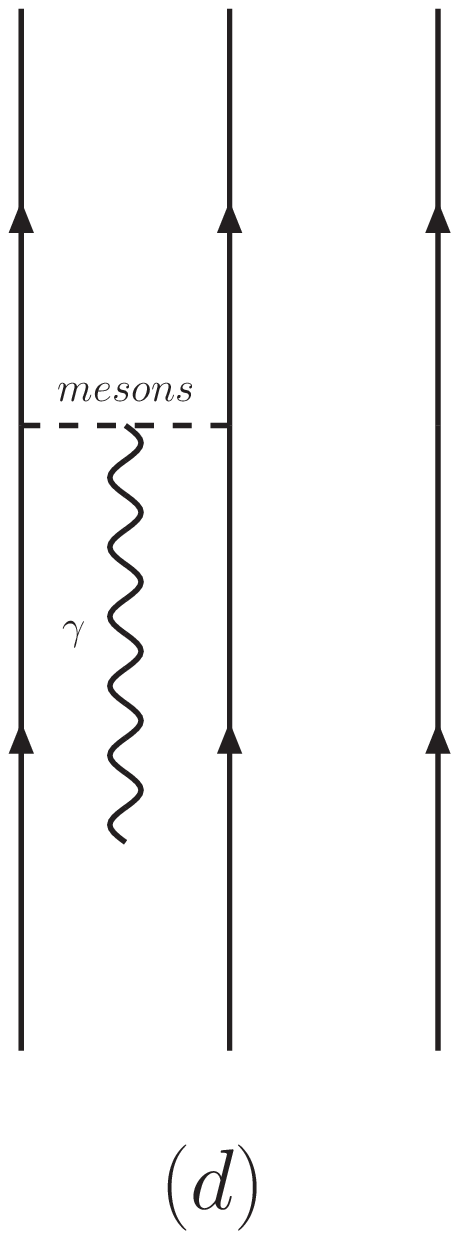}
  \includegraphics[bb=80 450 330 800,clip=true,scale=0.34]{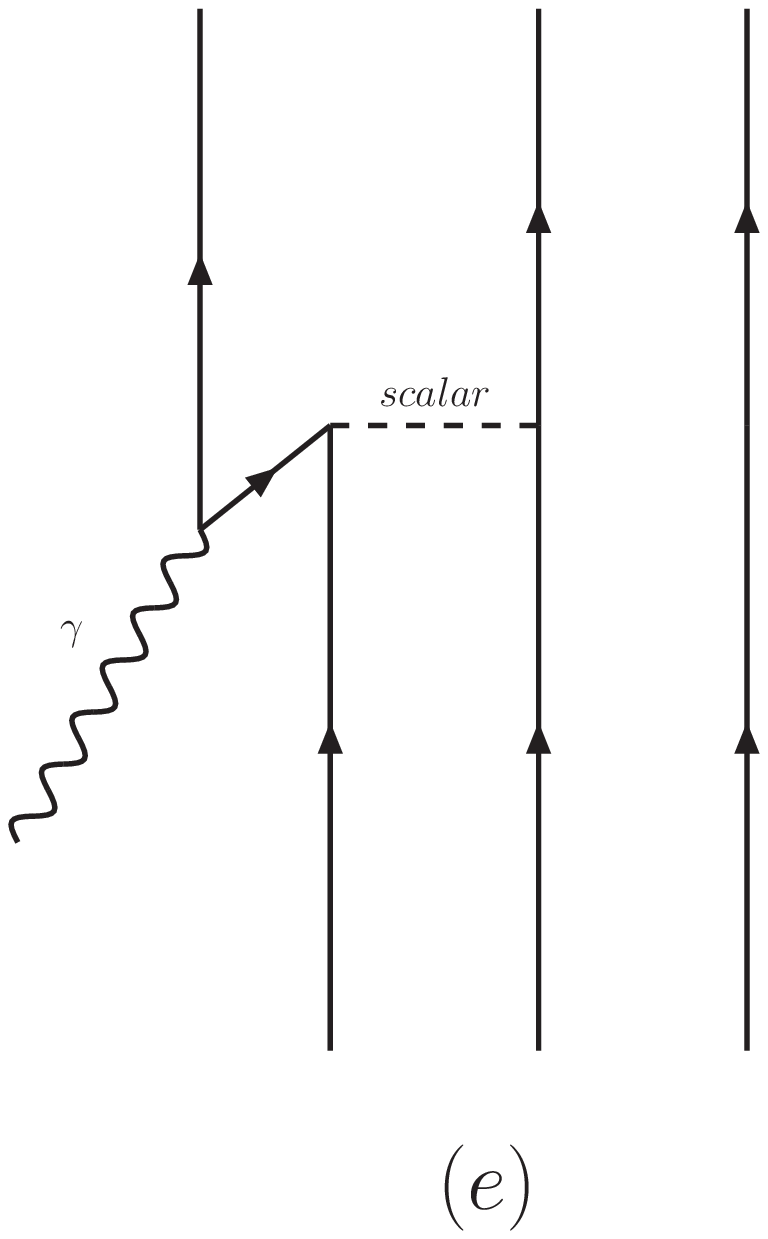}
 \caption{\small One-body and two-body exchange currents between
quarks: (a) impulse, (b) gluon pair, (c) meson ($\pi$, $\eta$,
$\eta'$, $\rho$, $\omega$) pair, (d) mesonic, (e)
scalar(confinement, $\sigma$) pair.}\end{figure*}

\textbf{a) charge density}
\begin{eqnarray}
\label{imp} {\rho}_{imp}^{IS/IV} ({\bf r}_i,{\bf q)} & = & {Q_i}e
e^{i{\bf{q \cdot r }}_i}
\end{eqnarray}

\begin{eqnarray}
{ \rho}_{gq{\bar q}} ({\bf r}_i,{\bf r}_j,{\bf q})& = &
-i{{\alpha_s}\over{16 m_q^3
}}\,{\bm{\lambda}^c_i}\cdot{\bm{\lambda}^c_j} \left \lbrace
{Q_i}ee^{i{\bf q}\cdot {\bf r}_i} \left [ {\bf q} \cdot {\bf r} +
({\bm {\sigma}_i}\times{\bf q}) \cdot ({\bm{\sigma}_j}\times{\bf r
}) \right ] +(i\leftrightarrow j) \right\rbrace {1\over
r^3},\hspace{10pt}\
\end{eqnarray}

\begin{eqnarray}
&&\rho_{\pi q{\bar q}}({\bf r}_i,{\bf r}_j,{\bf q}) =ee^{i{\bm
q}\cdot{\bm x}_1}(\frac{1}{3}{\bm \tau}^i\cdot{\bm
\tau}^j+\tau_{jz}) {\bm \sigma}_i\cdot\frac{i{\bm q}}{2m_q}{\bm
\sigma}_j\cdot {\bm
\nabla}_rf_\pi(r)+(i\leftrightarrow j),\\
&&\rho_{\eta q{\bar q}}({\bf r}_i,{\bf r}_j,{\bf q}) =ee^{i{\bm
q}\cdot{\bm x}_1'}(\frac{1}{3}+\tau_{iz}) {\bm
\sigma}_i\cdot\frac{i{\bm q}}{2m_q}{\bm \sigma}_j\cdot {\bm
\nabla}_rf_\eta(r)+(i\leftrightarrow j),\\
&&\rho_{\rho q{\bar q}}({\bf r}_i,{\bf r}_j,{\bf q})=ee^{i{\bm
q}\cdot{\bm x}_i'}(\frac{1}{3}{\bm \tau}^i\cdot{\bm
\tau}^j+\tau_{jz}) {\bm \sigma}_i\times\frac{i{\bm
q}}{2m_q}\cdot{\bm \sigma}_j\times {\bm
\nabla}_rf_\rho(r)+(i\leftrightarrow j),\\
&&\rho_{\omega q{\bar q}}({\bf r}_i,{\bf r}_j,{\bf q})=ee^{i{\bm
q}\cdot{\bm x}_i'}(\frac{1}{3}+\tau_{iz}) {\bm
\sigma}_i\times\frac{i{\bm q}}{2m_q}\cdot{\bm \sigma}_j\times {\bm
\nabla}_rf_\omega(r)+(i\leftrightarrow j).
\end{eqnarray}

\begin{eqnarray}
\rho_{\gamma\pi\pi}({\bf r}_i,{\bf r}_j,{\bf
q})\approx0,\hspace{20pt}\rho_{\gamma\rho\rho}({\bf r}_i,{\bf
r}_j,{\bf q})\approx0.
\end{eqnarray}

\begin{eqnarray}
\rho_{S}({\bf r}_i,{\bf r}_j,{\bf q}) &=& e^{i{\bf q}\cdot {\bf
r}_i}{Q_i}e {1\over (2m_q)^3}\left ( {3\over2} {\bf q}^2-i{\bf
q}\cdot {\bm \nabla}_r +{1\over 2}{\bm \nabla}_r^2 \right )
V_{S}({\bf r}_i,{\bf r}_j).
\end{eqnarray}

\textbf{b) exchange current}
\begin{eqnarray}
{\bm J}_{imp}({\bf r}_i,{\bf q})& = & {Q_i}e{{i}\over{2 m_q
}}(i[{\bm \sigma}_i\times{\bm p}_i,e^{i{\bf q}\cdot {\bf
r}_i}]+\{{\bm p}_i,e^{i{\bf q}\cdot {\bf r}_i}\}).
\end{eqnarray}

\begin{eqnarray}
{\bf J}_{gq{\bar q}} ({\bf r}_i,{\bf r}_j,{\bf q})& = &
-{{\alpha_s}\over{4 m_q^2
}}\,{\bm{\lambda}^c_i}\cdot{\bm{\lambda}^c_j} \Bigl\lbrace
{{Q_i}ee^{i{\bf q}\cdot {\bf r}_i}}{1\over 2} ({\bm
{\sigma}_i}+{\bm{\sigma}_j})\times{\bf r} +(i\leftrightarrow
j)\Bigr\rbrace {1\over r^3}.
\end{eqnarray}

\begin{eqnarray}
 {\bm
 j}^1_{\pi q{\bar q}}({\bf r}_i,{\bf r}_j,{\bf q})&=&ee^{i{\bm
 q}\cdot{\bm r}_1}(\frac{1}{3}{\bm \tau}_i\cdot{\bm \tau}_j+\tau_{jz})\frac{i{\bm q}}{4m_q^2}\times{\bm \nabla}_r{\bm \sigma}_j\cdot{\bm
 \nabla}_rf_\pi+(i\leftrightarrow
j),\\
{\bm j}^{2}_{\pi q{\bar q}}({\bf r}_i,{\bf r}_j,{\bf q})&
=&ee^{i{\bm q}\cdot{\bm x}_1}[{\bm \tau}^i\times{\bm
\tau}^j]_3{\bm \sigma}_i({\bm \sigma}_j\cdot{\bm
\nabla}_r)V_\pi+(i\leftrightarrow
j), \\
{\bm
 j}_{\eta q{\bar q}}({\bf r}_i,{\bf r}_j,{\bf q})&=&ee^{i{\bm
 q}\cdot{\bm r}_1}(\frac{1}{3}+\tau_{iz})\frac{i{\bm q}}{4m_q^2}\times{\bm \nabla}_r{\bm \sigma}_j\cdot{\bm
 \nabla}_rf_\eta+(i\leftrightarrow
j),\\
{\bm j}^{1}_{\rho q{\bar q}}({\bf r}_i,{\bf r}_j,{\bf q}) &=&
ee^{i{\bm q}\cdot{\bm x}_1}(\frac{1}{3}{\bm \tau}_i\cdot{\bm
\tau}_j+\tau_{iz})({\bm \sigma}_1+{\bm \sigma}_2)\times{\bm
\nabla}_rV_\rho+(i\leftrightarrow
j),\\
{\bm j}^{2}_{\rho q{\bar q}}({\bf r}_i,{\bf r}_j,{\bf q})&
=&ee^{i{\bm q}\cdot{\bm x}_1}[{\bm \tau}^i\times{\bm
\tau}^j]_3[{\bm \nabla}_r-{\bm \sigma}_1\times({\bm
\sigma}_2\times{\bm \nabla}_r]f_\rho+(i\leftrightarrow
j),\\
{\bm j}_{\omega q{\bar q}}({\bf r}_i,{\bf r}_j,{\bf q}) &=&e
e^{i{\bm q}\cdot{\bm x}_1}(\frac{1}{3}+\tau_{jz})({\bm
\sigma}_1+{\bm \sigma}_2)\times{\bm
\nabla}_rf_\omega+(i\leftrightarrow j).
\end{eqnarray}

\begin{eqnarray}
{\bm j}_{\gamma\pi\pi}({\bf r}_i,{\bf r}_j,{\bf
q})&=&\frac{eg^2}{16\pi m_q^2}[{\bm \tau}_1\times{\bm
\tau}_2]_z{\bm \sigma}_1\cdot{\bm \nabla}_1{\bm \sigma}_2\cdot{\bm
\nabla}_2\int dve^{i{\bm q}\cdot({\bm R}-v{\bm r})}({\bm
z}_\mu\frac{e^{-L_\mu r}}{ L_\mu r}-{\bm
z}_\Lambda\frac{e^{-L_\Lambda r}}{L_\Lambda
r}),\hspace{20pt}\ \\
{\bm j}_{\gamma\rho\rho}({\bf r}_i,{\bf r}_j,{\bf
q})&=&\frac{eg^2}{16\pi m_q^2}[{\bm \tau}_1\times{\bm
\tau}_2]_z \nonumber\\
&\cdot&[4m_q^2-{\bm \sigma}_1\times{\bm \nabla}_1\cdot {\bm
\sigma}_2\times{\bm \nabla}_2]\int dve^{i{\bm q}\cdot({\bm
R}-v{\bm r})}({\bm z}_\mu\frac{e^{-L_\mu r}}{L_\mu r }-{\bm
z}_\Lambda\frac{e^{-L_\Lambda r}}{L_\Lambda r}).\hspace{15pt}\
\end{eqnarray}

\begin{eqnarray}
{\bf J}^{IS/IV}_{S}({\bf r}_i,{\bf r}_j,{\bf q}) & = & -{1\over
2m_q^2} \Bigl \{ {Q_i}e e^{i {\bf q}\cdot {\bf r}_i} \bm{\sigma}_i
\times {\bf q}   V_{S}({\bf r}_i,{\bf r}_j) +(i\leftrightarrow j)
\Bigr \}.
\end{eqnarray}

We have used the following abbreviations
 ${\bf R}=
({\bf r}_i+{\bf r}_j)/2 $, ${\bm z}_m({\bf q},{\bf r})=L{\bf
r}+ivr{\bf q}$, and $L(q,v) = [{1\over 4}q^2(1-4v^2)+\mu^2]^{1/2}
$ in Eqs. (A17, A18).

In the above derivation, we use strong interaction form factors
given by Adam $et.\ al.$ \cite{Adam}, and neglect the nonlocal
terms as usual.

\end{document}